\documentclass[conference,anonymous=true]{IEEEtran}

\IEEEoverridecommandlockouts
\usepackage{cite}
\usepackage{amsmath,amssymb,amsfonts}
\usepackage{algorithmic}
\usepackage{graphicx}
\usepackage{textcomp}
\usepackage{xcolor}
\graphicspath{{Images/}}
\usepackage{enumitem}
\usepackage[colorlinks=true]{hyperref}

\usepackage{colortbl}  % For row colors
\usepackage{booktabs}

\usepackage{listings}
\setlist[itemize,enumerate]{left=3pt} % Decrease left margin to 10pt (default is more)
\setlist[itemize]{itemsep=1pt, topsep=2pt} % Adjust space between items and above the list

\def\BibTeX{{\rm B\kern-.05em{\sc i\kern-.025em b}\kern-.08em
    T\kern-.1667em\lower.7ex\hbox{E}\kern-.125emX}}

\begin{document}

\title{An LLM-Integrated Framework for Completion, Management, and Tracing of STPA}

% \author{\textit{Redacted for Double-Blind Review}}
\author{
	\IEEEauthorblockN{Ali Raeisdanaei$^1$, Juho Kim$^{1,2}$, Michael Liao$^{1,2}$, and Sparsh Kochhar$^{1,2}$}
	\IEEEauthorblockA{
		\textit{$^1$Blue Sky Solar Racing} \\
		\textit{$^2$University of Toronto} \\
		Toronto, Ontario, Canada \\
		ali.raeisdanaei@gmail.com \quad \{juho.kim,michael.liao,sparsh.kochhar\}@mail.utoronto.ca \\
	}
}

\maketitle

\begin{abstract}
	In many safety-critical engineering domains, hazard analysis techniques are an essential part of requirement elicitation.
	Of the methods proposed for this task, STPA (System-Theoretic Process Analysis) represents a relatively recent development in the field.
	The completion, management, and traceability of this hazard analysis technique present a time-consuming challenge to the requirements and safety engineers involved.
	In this paper, we introduce a free, open-source software framework to build STPA models with several automated workflows powered by large language models (LLMs).
	In past works, LLMs have been successfully integrated into a myriad of workflows across various fields.
	Here, we demonstrate that LLMs can be used to complete tasks associated with STPA with a high degree of accuracy, saving the time and effort of the human engineers involved.
	We experimentally validate our method on real-world STPA models built by requirement engineers and researchers.
	The source code of our software framework is available at the following link: \url{https://github.com/blueskysolarracing/stpa}.
\end{abstract}


\section{Introduction}

Historically, the growing complexity of safety-critical systems drove the development of more nuanced theories and applications for their safety analysis~\cite{checkland2000systems}.
Systematic safety analyses are required by safety compliance standards to elicit requirements~\cite{ISO26262}.
Among the methods proposed, System-Theoretic Process Analysis (STPA) is a relatively new hazard analysis technique, developed at the Massachusetts Institute of Technology (MIT), founded on System-Theoretic Accident Model and Processes (STAMP).
Leveson and Thomas~\cite{STPA_Handbook} note that STPA is not only more time and resource-efficient than the traditional offerings, but is also capable of finding more causal scenarios than those found by legacy methods~\cite{leveson_saferworld}.

The STPA Handbook~\cite{STPA_Handbook}, a comprehensive text on the technique authored by the pioneers of the framework, outlines the four steps involved in the hazard analysis process:

\begin{enumerate}
	\item \label{step:1} Losses and hazards are defined, with the process aimed at minimizing their occurrence. 
	\item \label{step:2} A model of controllers in the system is created to show the controlled actions (and feedback loops) of a system.
	\item \label{step:3} Unsafe Control Actions (UCAs) are elicited from the control structure (from the Step~\ref{step:2}) that lead to hazards in the first step.
	\item \label{step:4} Loss scenarios are considered as the causal factors that could lead a UCA to a hazard.
\end{enumerate}

The STPA artifacts generated through these steps can then be used to derive system requirements, each justifying their need by a concrete trace to a hazard and loss.

According to Leveson and Thomas~\cite{STPA_Handbook}, STPA is notable for its ability to identify software-related scenarios that traditional methods tend not to find.
In \textbf{our first contribution}, we seek to amplify this ability by creating a pure Python framework for constructing STPA artifacts that can be explicitly traced to requirements and design, as well as being tightly integrated within an existing software system.
To the best of our knowledge, this framework is the first of its kind where artifacts are programmatically integrated within the software system.
With that said, the use of our software is not limited exclusively to software-centric systems, as other systems can still benefit from several features offered by our library.

Gotel et al.~\cite{gotel_analysis_of_req_trace_prob} define requirement traceability as ``the ability to describe and follow the life of a requirement (i.e., from its origins, through its development and specification, to its subsequent deployment and use''~\cite{gotel_analysis_of_req_trace_prob}.
Managing the traceability between the origins of the requirements from STPA then to the software can be quite challenging and often missing~\cite{RE_as_success_factor_in_software_projects, improving_traceability, survery_traceability_in_re}.
This truism is especially more apparent in the commonly used tools for STPA management such as Microsoft Word or Excel~\cite{handbook_of_systems_thinking_methods}.
Our framework forces the users to express each artifact (losses, hazards, UCAs, etc.) while also defining its parts (source, type, links, and more) in Python.
Thus, the modeled system's STPA artifacts can reliably be accessed and analyzed as a concrete, traceable package with scripts in Python or other programming languages that support Python interoperability (it is also worth mentioning that Python is a widely popular scripting language).
This approach of defining STPA in a Pythonic idiom, also allows requirement engineers to benefit from standard software engineering practices and principles such as version control, type/style checks, and more.

One drawback of STPA is that the number of artifacts has the potential to grow exponentially.
UCAs are generated by considering each control action against four failure types -- while not always the case, very often all four types of UCAs are created for each control action.
Similarly, in turn, each UCA is considered for four different loss scenario types.
For example, in a small system of only 5 components and 20 control actions, 80 UCAs as well as 320 loss scenarios can potentially be generated, which can be very costly to produce and maintain for a team creating a small system of 5 components.
This scaling of the number of artifacts places a major limitation and challenge in the completion of STPA.

We believe (and past works~\cite{CHARALAMPIDOU2024106608, era_LLM_STPA} have shown) that the generation of UCAs and loss scenarios (Steps~\ref{step:3} and~\ref{step:4}) can be augmented with large language models (LLMs) to help human engineers speed up their workflow.
LLMs are advanced machine learning models trained on large amounts of data, enabling them to intake and generate human-like text. 
They have shown extensive capability when applied in various fields such as social science~\cite{kim-guerzhoy-2024-observing}, psychology~\cite{10.5555/3618408.3618425}, and game-theory~\cite{doi:10.1126/science.ade9097}.
As LLMs are often characterized by their strengths in pattern recognition, contextual reasoning, and structured text generation, we hypothesize they may be well suited for systematic, rule-based analyses such as STPA.
Furthermore, LLMs can process and structure large amounts of information at scale, making them particularly useful for handling the extensive artifact generation inherent in STPA.
By automating repetitive, yet critical, parts of the analysis, LLMs may reduce the cognitive load on safety engineers and help ensure consistency in hazard identification.

\textbf{Our second contribution} is the enhancement of our STPA framework with LLM integration to generate UCAs, loss scenarios, and links.
This could be especially effective because the domain of STPA is closed -- the model of the system being analyzed can be given as input (the control structure diagram, artifacts, etc.), and each generation is paired with a systematic prompt from the STPA Handbook.
In addition, the use of LLMs can be effective in bypassing the confirmation bias of the requirement engineer, which has been shown to miss many failures in system safety~\cite{taxonomy_of_fallacies_in_system_safety_arguments}.
LLMs have already been used in the space of requirement engineering, and they have been shown to be effective when answering questions such as ``What can go wrong?''~\cite{avoiding_confirmation_bias_in_safety_management_system}.
With that said, a requirement engineer in the loop can then judge the validity of the generated UCAs and loss scenarios.

We evaluate our LLM enhancement of STPA against a case study from the STPA Handbook~\cite{STPA_Handbook} with the following research questions:
\begin{enumerate}[label=RQ-\arabic*]
	\item \label{RQ:1} How many of the LLM-generated UCAs and loss scenarios will be selected by the requirement engineer as valid to the system and its environment?
	\item \label{RQ:2} How accurate is the LLM in linking a related artifact to another (e.g., a loss to a hazard)?
\end{enumerate}

\section{Background}\label{sec:background}

\subsection{Overview of STPA}
According to the STPA Handbook,

\begin{quotation}
	STPA (System-Theoretic Process Analysis) is a relatively new hazard analysis technique based on an extended model of accident causation.
	In addition to component failures, STPA assumes that accidents can also be caused by unsafe interactions of system components, none of which may have failed.~\cite{STPA_Handbook}
\end{quotation}

STPA analysis is carried out in four steps: \ref{step:1}) defining the purpose of the analysis, \ref{step:2}) modeling the control structure, \ref{step:3}) identifying UCAs, and \ref{step:4}) identifying loss scenarios.

Step~\ref{step:1} starts with defining losses which ``involve something of value to stakeholders.
Losses may include a loss of human life or human injury, property damage, environmental pollution, [\dots]''~\cite{STPA_Handbook}.
This step also defines hazards -- ``a system state or set of conditions that, together with a particular set of worst-case environmental conditions, will lead to a loss''~\cite{STPA_Handbook} -- and system constraints ``specifying system conditions or behaviors that need to be satisfied to prevent hazards''~\cite{STPA_Handbook}.

\begin{figure}[ht!]
	\centering
	\includegraphics[width=0.35\linewidth]{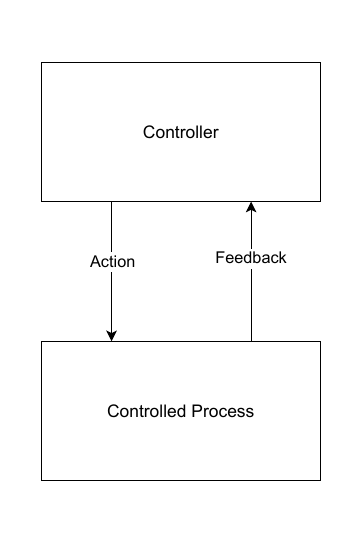}
	\caption{A diagram of a simple control structure. This \texttt{draw.io} diagram is a reproduction of Figure 2.6 in the STPA Handbook~\cite{STPA_Handbook}.}
	\label{fig:simple_control_structure}
\end{figure}

\begin{figure*}[ht!]
	\centering
	\includegraphics[width=0.6\linewidth]{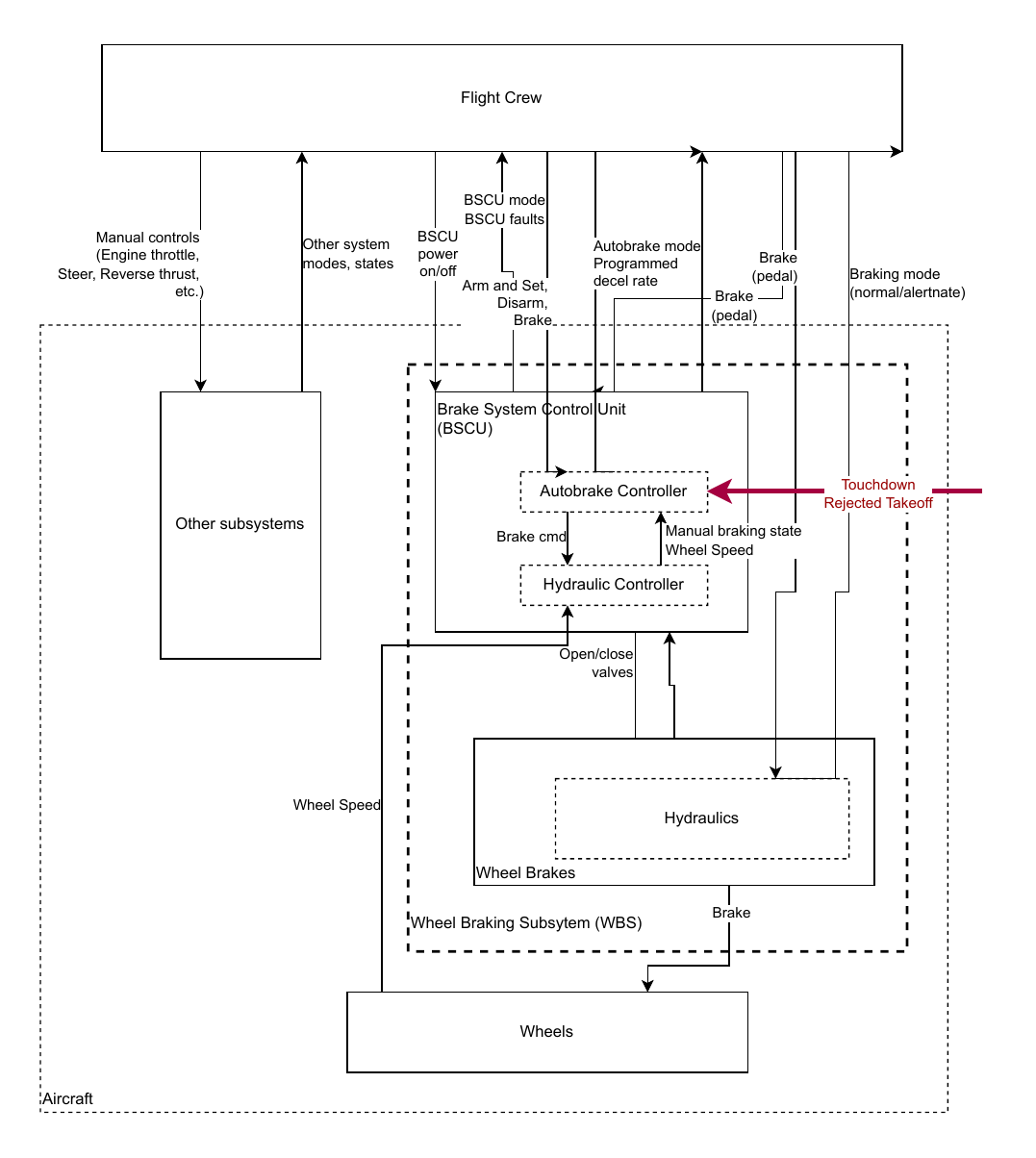}
	\caption{An example control structure of an aviation wheel braking system~\cite{STPA_Handbook, autohold_STPA_exemplar}. This \texttt{draw.io} diagram is a reproduction of Figure 2.12 in the STPA Handbook~\cite{STPA_Handbook}.}
	\label{fig:example}
\end{figure*}

In Step~\ref{step:2}, a ``hierarchical control structure, [\dots] a system model that is composed of feedback control loops''~\cite{STPA_Handbook}, is created as a diagram.
A control structure is modeled to show control and feedback of processes or components through \textit{control actions} and \textit{feedbacks}.
An example of a simple control structure is shown in Figure~\ref{fig:simple_control_structure}, and a realistic example from Chapter 2 of the handbook is shown in Figure~\ref{fig:example}.

UCAs are ``control actions that, in a particular context and worst-case environment, will lead to a hazard [\dots] There are four ways a control action can be unsafe:''~\cite{STPA_Handbook}

\begin{enumerate}
	\item It is not provided, leading to a hazard.
	\item It is provided, but leads to a hazard.
	\item It is provided too early/late or out-of-order, leading to a hazard.
	\item It is provided for too long or stopped prematurely, leading to a hazard.
\end{enumerate}

Each control action from the control structure model of the second step is considered for each of the above types when generating UCAs during Step~\ref{step:3}.

Lastly, in Step~\ref{step:4}, loss scenarios are generated.
``A loss scenario describes the causal factors that can lead to the UCAs and to hazards.''
Leveson describes two types of loss scenarios and two subtypes for each type (for a total of four subtypes).
While all subtypes lead to a hazard (and are explicitly linked as such), only two of them explicitly link to an improperly executed or unexecuted UCA.
The UCAs from Step~\ref{step:3} are considered for each of the four subtypes of loss scenarios.

\begin{figure}[ht!]
	\centering
	\includegraphics[width=\linewidth]{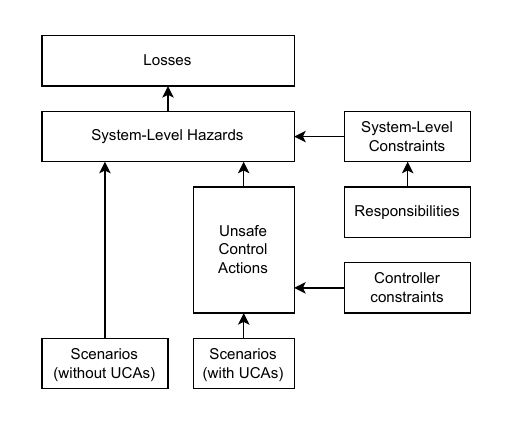}
	\caption{Outputs of the four steps of STPA showing their traceability as depicted in the handbook~\cite{STPA_Handbook} (Figure 2.21).}
	\label{fig:stpa_overview}
\end{figure}

The outputs of STPA are explicitly linked to each other during their construction so that all of them eventually lead to a loss or losses; this can be seen in Figure~\ref{fig:stpa_overview}, taken directly from the handbook~\cite{STPA_Handbook}.
Figure~\ref{fig:stpa_overview} also shows other artifacts such as system-level constraints and responsibilities; however, we do not focus on them for the scope of this paper.

\subsection{LLMs and Safety Engineering}

LLMs are artificial intelligence systems designed to process and generate human-like text and are capable of performing tasks such as text completion, summarization, code generation, decision support, and more.
Consisting of billions of parameters and trained with trillions of tokens, LLMs allowed various linguistic tasks to be automated with high efficiency in other fields~\cite{kim-guerzhoy-2024-observing}.

There exist various prompting techniques to accomplish a task using LLMs (chain-of-thought, zero-shot, few-shot, and more).
In this paper, we explored using few-shot prompting to generate UCAs and loss scenarios.
Few-shot prompting allows us to supply several example outputs to an LLM which helps guide it to generate artifacts that align with the existing distribution of definitions.

The use of natural language processing (NLP) and LLMs has been explored in the space of safety engineering~\cite{nlp_aviation_safety, nlp_aviation_safety2, nlp_accident_reports}.
More specifically, the use of LLMs on other hazard analysis techniques such as HAZOP has been studied~\cite{nlp_HAZOP, nlp_HAZOP_deviations}.
LLMs have also been readily used in assurance cases to generate defeater arguments; these are counter-arguments that challenge the safety engineers reasoning about a system~\cite{Viger_AI_Supported_EA, avoiding_confirmation_bias_in_safety_management_system, eval_eff_GPT_defeaters, CoDefeater_LLM_AC}.

Several prior works~\cite{LLM_in_HazardAnal, LLM_Supported_Safety_Engineering, era_LLM_STPA, CHARALAMPIDOU2024106608} propose some forms of using LLMs for generating STPA outputs.
Following evaluations, Diemert and Weber~\cite{LLM_in_HazardAnal} and Stravroula et al.~\cite{CHARALAMPIDOU2024106608} found correctness of outputs to be less than half or about half, respectively, Hong et al.~\cite{LLM_Supported_Safety_Engineering} did not perform a large-scale survey of result qualities but nonetheless found hazards ``worth addressing'', while Qi et al.~\cite{era_LLM_STPA} found that results are unreliable without human interventions.

Our approach to using LLMs for STPA differs significantly and offers several novel aspects compared to prior works.
First, we pass an image of the control structure diagram with our prompts, making full use of OpenAI's latest vision model: GPT-4o~\cite{openai2024gpt4technicalreport}.
Previous approaches~\cite{era_LLM_STPA, CHARALAMPIDOU2024106608} used ChatGPT with the older GPT-4~\cite{openai2024gpt4technicalreport} model that did not support image inputs.
The incorporation of the visual representation of the system offers a distinct advantage by providing the LLM with the context of the entire system (i.e., entities, control actions, feedbacks, and associated labels) during output generation whereas a previous approach~\cite{CHARALAMPIDOU2024106608} only supplied a single control action as part of the prompt.
This limitation prevented the LLM from taking a holistic view of the system during synthesis.

Second, we also pass the entire STPA context (losses, hazards, and, for loss scenario generation, UCAs) in addition to the control structure diagram as part of the prompt to allow the LLM to directly link losses and hazards it finds appropriate for each output generated. Unlike in the previous approach~\cite{CHARALAMPIDOU2024106608}, our prompt synthesizes multiple outputs in a single query. Previously, each query only outputs a single definition. Our approach is, therefore, more efficient by at least an order of magnitude in terms of the number of prompts (we generate 50 UCAs/loss scenarios per prompt in our experiments).
This approach also ensures that the generated outputs are sufficiently diverse, as the LLM has context of the previous ones it generated.

Third, our LLM-powered methods can directly be integrated with existing utilities and tools available in our software library, and can easily be extended or modified to suit the user's needs.

Several meaningful advancements in LLM capabilities in addition to the support for vision were introduced over the years since GPT-4~\cite{openai2024gpt4technicalreport}'s inception. We believe this makes our exploration with the latest offering timely.

\subsection{STPA Ecosystem}

There is a surprising lack of tools or frameworks in the STPA ecosystem, largely owing to the method's recency.
With that said, there does exist an Eclipse-based STPA verifier plugin\footnote{The STPA verifier plugin is part of a larger framework dedicated to STAMP-related tools. It is available at the following link: \url{https://github.com/SE-Stuttgart/XSTAMPP}} for managing STPA~\cite{eclipse_plugin_STPA_1, eclipse_plugin_STPA_2}, developed at the University of Stuttgart, and demonstrated in a case study involving BMW~\cite{industrial_case_study_stpa}.
Their plugin has also been extended with a formal model checker to verify the software implementation according to the STPA results~\cite{eclipse_plugin_STPA_verification}.
The integration of our framework and their plugin tool may yield interesting avenues for future research and application of STPA.

\subsection{Regulatory Compliance}

Hillen et al.~\cite{HARA_LLM} build on the work by Diemert and Weber~\cite{LLM_in_HazardAnal} and introduces LASAR which supports Hazard Assessment \& Risk Analysis (HARA) of automotive systems under ISO~26262-3.6 \cite{ISO26262}.
LASAR is especially useful because it cooperates with the safety engineer in providing severity, exposure, and controllability ratings \emph{with reasoning behind its results}.
They evaluate their tool on a small case study and report its usefulness according the safety engineers.
LASAR takes a ``situation catalogue'' from ISO~26262-2~\cite{ISO26262} which might be a predetermined internal list or, by the recommendations of ISO~26262-2, generated from a systema
tic hazard analysis technique such as FMEA or STPA:

\begin{quotation}
       \textbf{6.4.2.1} The operational situations and operating modes in which an item's malfunctioning behaviour
       will result in a hazardous event shall be described \dots \\
       \textbf{6.4.2.2} The hazards shall be determined systematically \dots \\
       \textbf{NOTE 1} FMEA approaches and HAZOP are suitable to support hazard identification \dots~\cite{ISO26262}
\end{quotation}

Thus, the outputs generated or artifacts stored by our framework may serve as crucial inputs to LASAR~\cite{HARA_LLM}.
However, we remain skeptical of the general ability of LLMs to \emph{reason} as they do in LASAR~\cite{HARA_LLM} about things like severity.
Our approach follows a systematic prompt following STPA guidelines which requires negative reasoning in the form of ``what can go wrong'' which we believe is especially more suited to a black box prone to hallucinations, viz. LLMs.
Note that reasoning about something like controllability is the opposite form of ``negative'' reasoning.

\section{Framework for Managing and Tracing STPA}

To effectively manage the complex artifacts of STPA and integrate them into software engineering workflows, we have developed a structured framework that models the STPA artifacts using object-oriented programming in Python.
In this framework, each concept in STPA -- for example, losses, hazards, UCAs, and loss scenarios -- is represented as a class.
This makes it easier to organize, trace, and manipulate these artifacts within the code.
According to the handbook, a hazard is structured with the following parts~\cite{STPA_Handbook}:
\begin{verbatim}
    <Hazard specification> =
    <System> &
    <Unsafe Condition> &
    <Link to Losses>
\end{verbatim}

A simplified version of the definition of hazard in our framework is shown in Figure~\ref{fig:hazard-class}.

\begin{figure}[ht!]
	\centering
	\begin{lstlisting}[language=Python]
from dataclasses import dataclass

@dataclass
class Hazard(Definition):
    system: str
    unsafe_condition: str
    losses: list[Loss]
	\end{lstlisting}
	\caption{The class definition for hazards. It contains three attributes: system, unsafe condition, and losses. The corresponding types are annotated next to each attribute name.}
	\label{fig:hazard-class}
\end{figure}

Note that different definitions in STPA make explicit connections to other artifacts.
For example, each hazard is linked to UCAs, which are, in turn, linked to loss scenarios.
Our design models this using \textit{has-a} relationships which can be seen in the \texttt{losses} attribute of the Python definition above.
This is how we preserve these relationships using explicit object references.

\begin{figure*}[ht!]
	\centering
	\includegraphics[width=\linewidth]{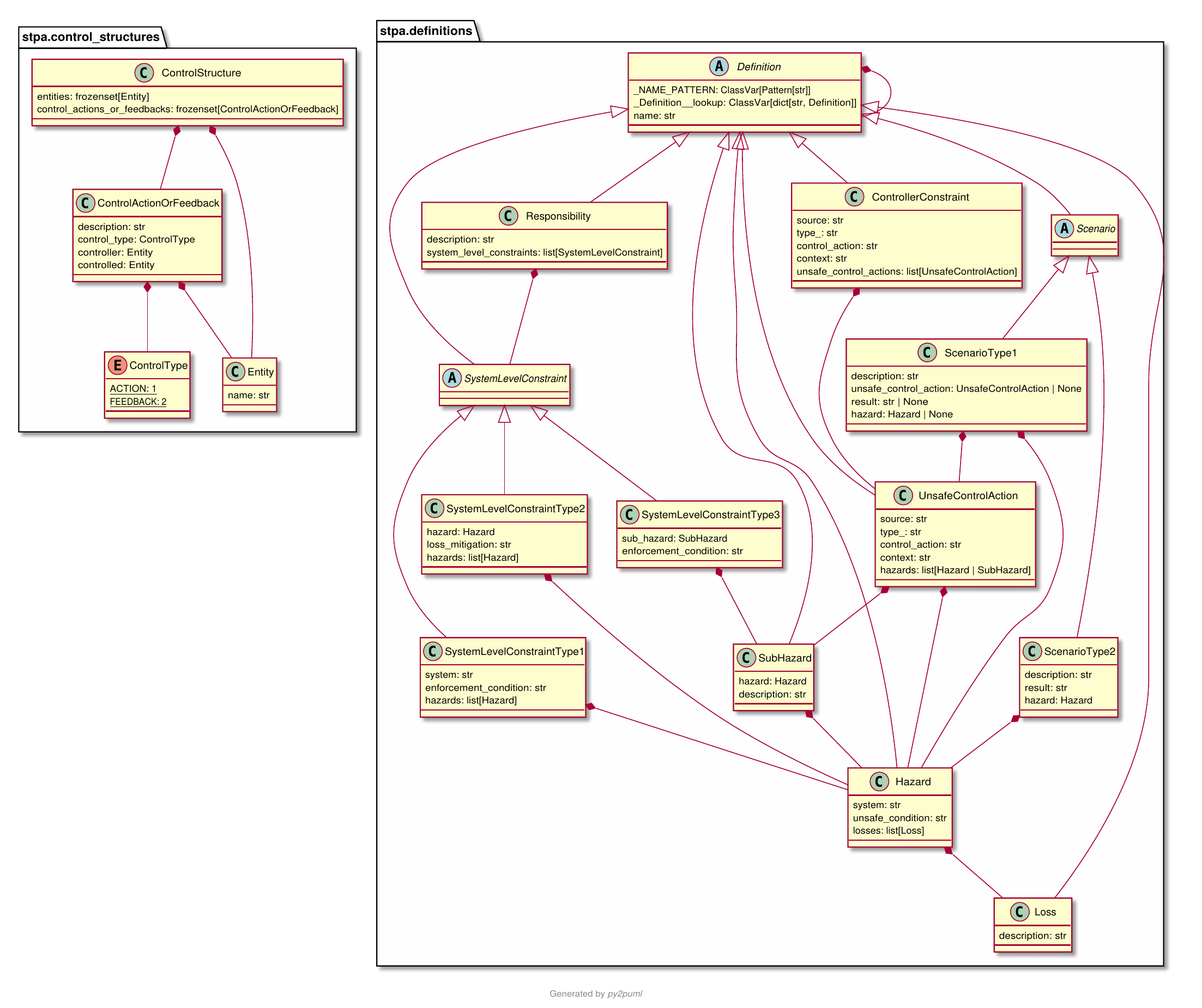}
	\caption{The UML class diagram for our STPA framework. The \texttt{stpa.control\_structures} section shows classes that pertain to elements within control structure diagram while the \texttt{stpa.definitions} section shows classes related to artifacts shown in Figure~\ref{fig:stpa_overview} plus several abstract classes.}
	\label{fig:stpa_class_diagram}
\end{figure*}

Figure~\ref{fig:stpa_class_diagram} shows the UML class diagram of our framework which gives an overview of all the defined classes, each representing a box in Figure~\ref{fig:stpa_overview}.

As previously mentioned, our framework can be integrated within a software system at the code level.
This enables developers and engineers to explicitly trace a software component to its requirement or to any other artifacts from STPA.
This is particularly important for maintaining requirement traceability, ensuring that safety constraints and requirements derived from STPA are not only documented but actively linked to the system's design and implementation.
For example, a software component may be designed to have an STPA \textit{Responsibility} in its object definition, as in Figure~\ref{fig:example-usage}.

\begin{figure}[ht!]
	\centering
	\begin{lstlisting}[language=Python]
from stpa.definitions import Responsibility

class Component:
  def __init__(self, ...,
      responsibilities: list[Responsibility]):
    ...
    assert len(responsibilities) > 0
    assert all(isinstance(r, Responsibility)
      for r in responsibilities)
    self.responsibilities = responsibilities
	\end{lstlisting}
	\caption{An example usage of our framework within existing code. Responsibilities are directly encoded within a component instance which may aid in requirement traceability during runtime.}
	\label{fig:example-usage}
\end{figure}

Having the requirements and elicitation with STPA being tightly integrated into software development is very useful for managing requirement traceability.
Our framework allows the access of STPA artifacts and requirements to the whole engineering team throughout the design, development, and verification stages.
Software engineering principles can be applied to the management and verification of the STPA development such as using version control systems like \texttt{git}, or easily generating models of the software along with its requirements and STPA artifacts.
For instance, all the losses and hazards of a system can be consolidated into a single ``loss-hazard'' file, as shown in Figure~\ref{fig:example-definitions} with a nuclear power plant example from Appendix A of the handbook~\cite{STPA_Handbook}.

\begin{figure}[ht!]
	\centering
	\begin{lstlisting}[language=Python, showstringspaces=false]
from stpa.definitions import (
  Definition, Hazard, Loss)

LOSSES = (
  Loss("L1", "People injured or killed"),
  Loss("L2", "Environment contaminated"),
  Loss("L3",
    "Equipment damage (economic loss)"),
  Loss("L4",
    "Loss of electrical power generation"),
)

HAZARDS = (
  Hazard("H1", "",
    "Release of radioactive materials",
    Definition.get_all("L1", "L2", "L3",
      "L4")),
  Hazard("H2", "Reactor",
    "temperature too high",
    Definition.get_all("L1", "L2", "L3",
      "L4")),
  Hazard("H3", "Equipment",
    "operated beyond limits",
    Definition.get_all("L3", "L4")),
  Hazard("H4", "Reactor", "shut down",
    Definition.get_all("L4")),
)
	\end{lstlisting}
	\caption{An example loss-hazard definitions for a nuclear power plant example from Appendix A of the handbook~\cite{STPA_Handbook}.}
	\label{fig:example-definitions}
\end{figure}

In our framework, we provide the following examples of system definitions with STPA introduced throughout the handbook~\cite{STPA_Handbook} (along with the control structure diagrams where applicable):

\begin{itemize}
	\item ``Nuclear Power Plant''~\cite{STPA_Handbook} (Appendix A)
	\item ``Aircraft''~\cite{STPA_Handbook} (Appendix A)
	\item ``Radiation Therapy''~\cite{STPA_Handbook} (Appendix A)
	\item ``Military Aviation''~\cite{STPA_Handbook} (Appendix A)
	\item ``Automotive''~\cite{STPA_Handbook} (Appendix A)
	\item ``Autonomous Auto Hold System''~\cite{STPA_Handbook} (Chapter 2, Appendix B, Appendix C)
	\item ``Autonomous H-II Transfer Vehicle''~\cite{STPA_Handbook} (Appendix B, Appendix C)
	\item ``Aircraft Flight Crew related to the Wheel Braking System''~\cite{STPA_Handbook} (Appendix C)
	\item ``Aircraft Brake System Control Unit (BSCU) Autobrake''~\cite{STPA_Handbook} (Appendix C)
	\item The full auto-hold system.~\cite{autohold_STPA_exemplar} (the one in the handbook is a modernized excerpt of this work).
\end{itemize}

The control structure model is to be generated using a popular open-source web-based flow-chart/online diagram creation software named \texttt{draw.io}\footnote{The \texttt{draw.io} web application is available on the following link: \url{https://draw.io/}} whose outputs are formatted in an open standard that derives from \texttt{XML}.
A parser implementation in our software library is capable of discerning, extracting, and instantiating control structures and their control actions and feedback loops as instances of relevant object classes.

These exemplars demonstrate how one may define and integrate STPA within existing projects.
Note that the use of our software is not strictly limited to software systems.
Indeed, on the contrary, any safety-critical system can be modeled using our framework, as demonstrated in several example system definitions we provide.

\section{Enhancing our STPA Framework with LLMs}

\subsection{UCA and Loss Scenario Syntheses}

\begin{figure*}[ht!]
	\centering
	\includegraphics[width=0.9\linewidth]{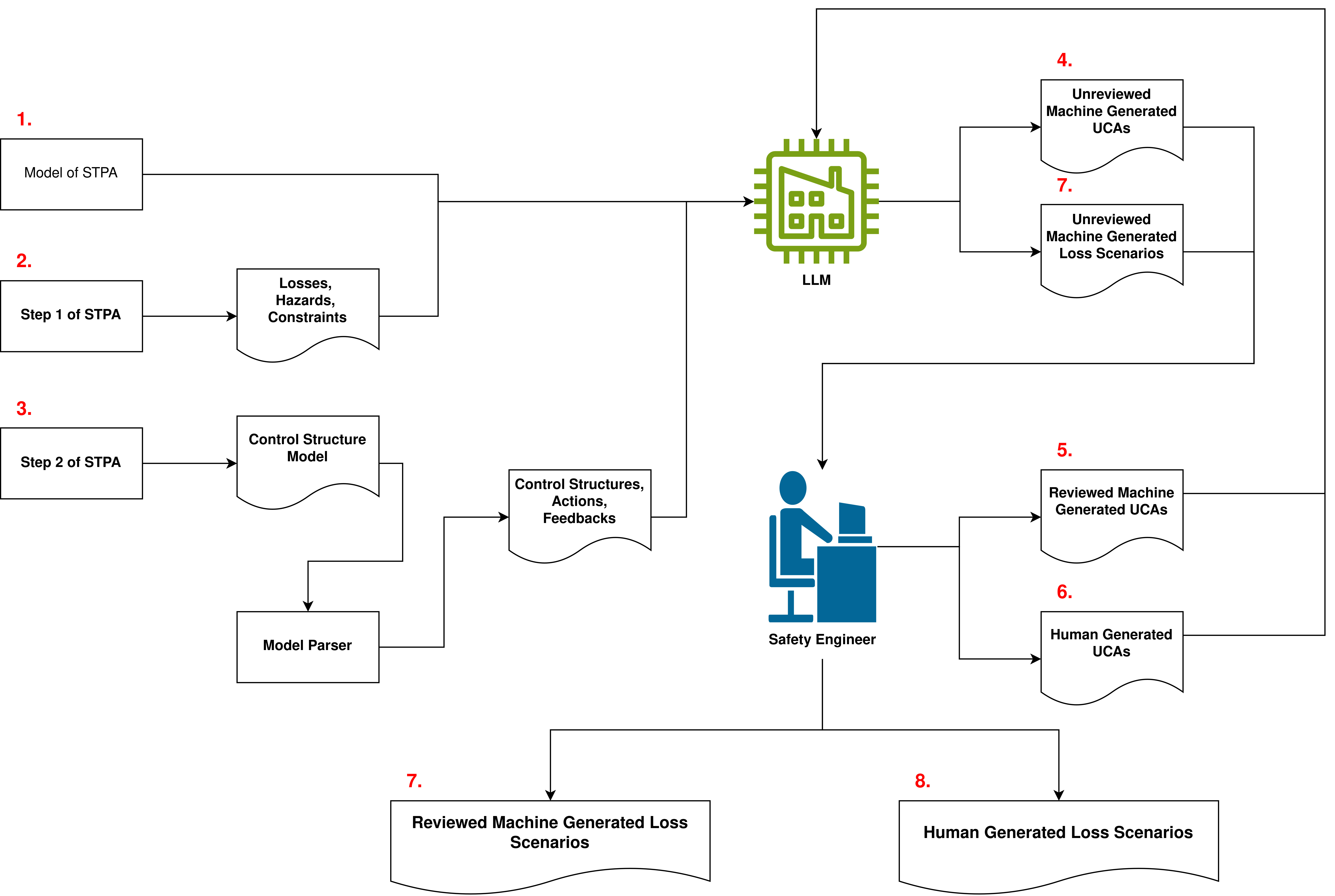}
	\caption{
		This diagram depicts the process that we use to enhance STPA with LLM.
		The model of the STPA along with the first two steps of it are given as input.
		The LLM is prompted to first produce a list of UCAs.
		The requirement engineer reviews and adds their own UCAs prompting the LLM to produce a list of loss scenarios.
		The loss scenarios are then reviewed by the requirement engineer and outputted along with loss scenarios generated by the requirement engineer.
		The inputs and outputs are labeled with their orders in red.
	}
	\label{fig:process}
\end{figure*}

We propose the following process depicted in Figure~\ref{fig:process} for the enhancement of our STPA framework with LLMs.
Steps~\ref{step:1} and~\ref{step:2} are delegated to the human engineers, as before, from which losses, hazards, and control structure diagrams are created (see Section~\ref{sec:background}).
These are then used as inputs to prompt the LLM to produce an arbitrary number of UCAs or loss scenarios (the number is passed as an argument and embedded within the prompt itself).
The prompts also contain a brief description of UCAs or loss scenarios.
As shown in the diagram, a requirement engineer oversees this process, reviews the generated UCAs and loss scenarios, and dismisses incorrect ones.
We define incorrect UCAs and incorrect loss scenarios as those that do not pertain to the system, its environment, or are reflective of LLM hallucinations.
The requirement engineer also adds their own UCAs and loss scenarios according to the STPA process as normal.

Note that it is also possible to use LLMs to produce other artifacts of STPA following the same workflow.
Just like for UCAs and loss scenarios, the base inputs of losses, hazards, as well as the model of the system can be passed as part of the LLM prompt, from which desired artifacts can be synthesized.
We have focused on just the UCAs and loss scenarios for the scope of this paper which is where, as we have previously noted, an explosion in the number of artifacts considered occurs.

We evaluate the LLM enhancement of our STPA framework using a case study in the STPA Handbook~\cite{STPA_Handbook}.

\subsection{Automated Linking of Artifacts}

While not the main focus of this paper, we also explore the use of LLMs for uses other than purely generative ones, namely testing if a definition should be linked to another. For each pair of possible linkages, we prompt an LLM to check if the two should be recommended for linkages. By suggesting possible linkages that can be made between definitions, a human engineer can save time looking up related definitions or find missing links that were previously overlooked which would enhance the traceability of the overall system.

\section{Case Studies of Systems from the STPA Handbook~\cite{STPA_Handbook}}

In our experiments, we used examples in the Handbook's~\cite{STPA_Handbook} Chapter 2 and Appendix A.

\subsection{Preliminary UCA and Loss Scenario Syntheses}

Chapter 2 contains definitions mostly pertaining to an aviation wheel braking system, the control structure of which can be seen in Figure~\ref{fig:example}.
There are some losses, hazards, and other STPA artifacts that are given in Chapter 2 of the handbook as examples from other systems, we later refer to these as non-canonical losses or hazards.
In total there are 8 losses and 8 hazards:

\begin{enumerate}[label=L-\arabic*]
	\item ``Loss of life or injury to people''~\cite{STPA_Handbook}
	\item ``Loss of or damage to vehicle''~\cite{STPA_Handbook}
	\item ``Loss of or damage to objects outside the vehicle''~\cite{STPA_Handbook}
	\item ``Loss of mission (e.g. transportation mission, surveillance mission, scientific mission, defense mission, etc.)''~\cite{STPA_Handbook}
	\item ``Loss of customer satisfaction''~\cite{STPA_Handbook}
	\item ``Loss of sensitive information''~\cite{STPA_Handbook}
	\item ``Environmental loss''~\cite{STPA_Handbook}
	\item ``Loss of power generation''~\cite{STPA_Handbook}
\end{enumerate}

\begin{enumerate}[label=H-\arabic*]
	\item ``Aircraft violate minimum separation standards in flight [L-1, L-2, L-4, L-5]''~\cite{STPA_Handbook}
	\item ``Aircraft airframe integrity is lost [L-1, L-2, L-4, L-5]''~\cite{STPA_Handbook}
	\item ``Aircraft leaves designated taxiway, runway, or apron on ground [L-1, L-2, L-5]''~\cite{STPA_Handbook}
	\item ``Aircraft comes too close to other objects on the ground [L-1, L-2, L-5]''~\cite{STPA_Handbook}
	\item \label{hazard:5} ``Satellite is unable to collect scientific data [L-4]''~\cite{STPA_Handbook}
	\item ``Vehicle does not maintain safe distance from terrain and other obstacles [L-1, L-2, L-3, L-4]''~\cite{STPA_Handbook}
	\item ``UAV does not complete surveillance mission [L-4]''~\cite{STPA_Handbook}
	\item \label{hazard:8} ``Nuclear power plant releases dangerous materials [L-1, L-4, L-7, L-8]''~\cite{STPA_Handbook}
\end{enumerate}

We supplied the losses and hazards to the following prompt to generate 50 UCAs.
The control structure of Figure~\ref{fig:example} was given in addition to the prompt as an image.
The prompt shown in Figure~\ref{fig:uca-prompt} was used to generate unsafe control actions.
Placeholder values, denoted with curly braces, are replaced with corresponding values during LLM invocation.

\begin{figure}[ht!]
	\centering
	\begin{lstlisting}[breaklines=true,postbreak=\mbox{\textcolor{red}{$\hookrightarrow$}\space},language=]
Generate {Number of UCAs to generate} unsafe control actions of the system shown in the control structure diagram.

An unsafe control action is linked to a single (sub)hazard and follows one of the four patterns:
 1) not providing causes hazard 
 2) providing causes hazard
 3) too early, too late, out of order or
 4) stopped too soon, applied too long.

An unsafe control action must contain the following five parts: <Source> <Type> <Control Action> <Context> <Link to Hazards>

Losses:

{Losses}

Hazards:

{Hazards}

Example unsafe control actions:

{Few-shot UCA examples}

Write each unsafe control action in a single line, preceded by a number.
	\end{lstlisting}
	\caption{The prompt passed to the LLM, in addition to the control structure diagram image, for generating UCAs. It contains the system's losses, hazards, and example UCAs (as few-shot examples). The prompt also includes a short description of UCAs and the number of UCAs that should be generated.}
	\label{fig:uca-prompt}
\end{figure}

As shown, the losses and hazards, along with unsafe control actions and (optionally) loss scenarios provided were supplied in our prompts as context and few-shot examples.
Similarly, we used the prompt shown in Figure~\ref{fig:loss-scenario-prompt} with the control structure diagram in Figure~\ref{fig:example} as an image to generate 50 loss scenarios.

\begin{figure}[ht!]
	\centering
	\begin{lstlisting}[breaklines=true,postbreak=\mbox{\textcolor{red}{$\hookrightarrow$}\space},numbers=right,language=]
Generate {Number of loss scenarios to generate} loss scenarios of the system shown in the control structure diagram.

A scenario must link to a single hazard. It may optionally also link to an unsafe control action.

Losses:

{Losses}

Hazards:

{Hazards}

Unsafe control actions:

{UCAs}

Example loss scenarios:

{Few-shot loss scenario examples}

Write each unsafe control action in a single line, preceded by a number.
	\end{lstlisting}
	\caption{The prompt passed to the LLM, in addition to the control structure diagram image, for generating loss scenarios. It contains the system's losses, hazards, UCAs, and example loss scenarios (as few-shot examples). The prompt also includes a short description of loss scenarios and the number of loss scenarios that should be generated.}
	\label{fig:loss-scenario-prompt}
\end{figure}

The outputs were all shuffled and exported to be analyzed for correctness by a group of human annotators familiar with STPA.
The annotations were one of \texttt{"CORRECT\_AND\_USEFUL"}, \texttt{"CORRECT\_BUT\_USELESS"}, and \texttt{"INCORRECT"}, as done by Diemert and Weber~\cite{LLM_in_HazardAnal} (the only difference in the labels are that they named a label as ``correct but not useful'' instead of our ``CORRECT\_BUT\_USELESS'').
These labels allow us to differentiate between annotations that were correct but not necessarily useful from a practical perspective.

The results are tabulated in Table~\ref{tab:chapter-2-evaluations}.
In our experiment, of the 50 generated UCAs, 39 were classified as \texttt{"CORRECT\_AND\_USEFUL"}, 6 as \texttt{"CORRECT\_BUT\_USELESS"}, and 5 as \texttt{"INCORRECT"}, yielding us an 78\% success rate for UCA synthesis.
For the 50 generated loss scenarios, surprisingly, 49 were classified as \texttt{"CORRECT\_AND\_USEFUL"} while just 1 was classified as \texttt{"CORRECT\_BUT\_USELESS"}.
No synthesized loss scenario was classified as \texttt{"INCORRECT"}.
This gave us a 98\% success rate for loss scenario synthesis.

\begin{table}[ht!]
	\centering
	\renewcommand{\arraystretch}{1.4}
	\setlength{\tabcolsep}{3pt}
	\caption{
		Number of human annotations for the UCAs and loss scenarios generated with an LLM for the auto-hold module from Chapter 2 of the STPA Handbook~\cite{STPA_Handbook}.
		78\% of the LLM generated UCA was labeled correct and useful, while 98\% of loss scenarios were labeled as correct and useful.
	}
	\label{tab:chapter-2-evaluations}
	\begin{tabular}{|l|c|c|c|}
		\hline
		\rowcolor{gray!30}
		\textbf{Artifact}      & \textbf{Correct and Useful} & \textbf{Correct but Useless} & \textbf{Incorrect} \\
		\hline
		\textbf{UCA}           & 39 (78\%)                   & 6 (12\%)                     & 5 (10\%)           \\
		\hline
		\textbf{Loss Scenario} & 49 (98\%)                   & 1 (2\%)                      & 0 (0\%)            \\
		\hline
	\end{tabular}
\end{table}

The first five output unsafe control actions for this example are listed below.

\begin{enumerate}[label=UCA-\arabic*]
	\item Autobrake Controller does not provide Brake command during emergency landing [H-3]
	\item Autobrake Controller provides Brake command during takeoff roll [H-3]
	\item Autobrake Controller provides Brake command too late after a rejected takeoff [H-3]
	\item Autobrake Controller stops providing Brake command too early during landing [H-3]
	\item BSCU does not engage Autobrake mode when required [H-1]
\end{enumerate}

Similarly, the first five loss scenarios for this exemplar are listed below.

\begin{enumerate}[label=Scenario \arabic*]
	\item Aircraft autopilot malfunctions mid-flight, leading to loss of minimum separation standards. [H-1]
	\item Structural failure of aircraft wing results in loss of airframe integrity. [H-2]
	\item Aircraft taxiing system error causes deviation from designated taxiway. [H-3]
	\item Ground proximity system failure leads to aircraft nearing other ground objects. [H-4]
	\item \label{loss-scenario:5} Satellite communication error prevents scientific data collection. [H-5]
\end{enumerate}

\subsection{Automated Linking of Artifacts}

For the experiment involving automated linking of artifacts, we used the definitions of losses and hazards in Appendix A of the STPA Handbook~\cite{STPA_Handbook}.
The hazards and losses of the five systems used in the experiment are sufficiently diverse in topic.
In total, 75 hazard-loss pairs were tested for linkages and compliance with the Handbook's annotations.

The results are tabulated in Table~\ref{tab:appendix-a-evaluations}.
Overall, results for four of the five systems exhibit high F-1 scores (88.0\%, 100.0\%, 95.0\%, and 100.0\%), respectively, but are noticeably lower for the radiation therapy example (66.7\%).
Note that the recall score for the radiation therapy is still quite high (100.0\%), suggesting that the LLM is outputting too many false positives.

Suspecting that poor data quality may be impacting the performance, we investigated deeper into the false positives that were encountered.
Upon taking a closer look at the handbook examples, we found several loss-hazard pairs that can definitely be linked together but have not been annotated as such in the handbook.
For example, Hazard ``H2: A nonpatient is unnecessarily exposed to radiation [L2]''~\cite{STPA_Handbook} should definitely be linked with Loss ``L4: Physical injury to a patient or nonpatient during treatment.''~\cite{STPA_Handbook}; however, the two have not been linked by the authors of the handbook.
This discovery further demonstrates that our method is capable of finding linkages missed even by the experts of STPA who annotated the handbook's exemplars.

\begin{table*}[ht!]
	\centering
	\renewcommand{\arraystretch}{1.4}
	\setlength{\tabcolsep}{3pt}
	\caption{Statistics on the comparisons between linkages generated by an LLM and annotations on the handbook~\cite{STPA_Handbook}.}
	\label{tab:appendix-a-evaluations}
	\begin{tabular}{|l|c|c|c|c|c|}
		\hline
		\rowcolor{gray!30}
		\textbf{System}              & \textbf{\# Pairs} & \textbf{Accuracy (\%)} & \textbf{Precision (\%)} & \textbf{Recall (\%)} & \textbf{F-1 Score (\%)} \\
		\hline
		\textbf{Nuclear Power Plant} & 16                & 81.3                   & 78.6                    & 100.0                & 88.0                    \\
		\hline
		\textbf{Aircraft}            & 14                & 100.0                  & 100.0                   & 100.0                & 100.0                   \\
		\hline
		\textbf{Radiation Therapy}   & 16                & 75.0                   & 50.0                    & 100.0                & 66.7                    \\
		\hline
		\textbf{Military Aviation}   & 21                & 90.4                   & 90.4                    & 100.0                & 95.0                    \\
		\hline
		\textbf{Automotive}          & 8                 & 100.0                  & 100.0                   & 100.0                & 100.0                   \\
		\hline
	\end{tabular}
\end{table*}

\section{Discussion}

\subsection{Significance of Evaluation Results}

Overall, our results show a very strong performance of LLMs for UCA and loss scenario generation and automatic linkages.
While incorrect or correct but useless outputs were indeed generated, their number remains small and we believe they will not detract from the benefits offered by having a large number of generated UCAs or loss scenarios available for the requirement engineer to select from.
In addition, we have demonstrated that our automated linkage annotator is capable of finding links that were otherwise missed even by the authors of the STPA Handbook~\cite{STPA_Handbook}.

The presence of a hazard unrelated to the control structure diagram led to an interesting observation.
Because the UCAs, by definition, are linked to a control action found in Figure~\ref{fig:example}, all the UCAs pertained to the aviation wheel braking system.
The few-shot examples may also have helped guide the LLM to do so.
As a result, Hazard~\ref{hazard:8} pertaining to a nuclear power plant was not used at all, as it did not relate to any control action in the diagram.
This suggests that, here, LLM was able to understand the system well enough to attribute the correct hazards and system processes.

Conversely, we found that 22 out of the 50 loss scenarios (44\%) generated directly linked to non-canonical hazards such as Hazard~\ref{hazard:5}.
An example of a non-canonical loss scenario is shown as~\ref{loss-scenario:5}.
With that said, all 22 of the loss scenarios were annotated as \texttt{"CORRECT\_AND\_USEFUL"}.
The fact that loss scenarios link to a non-canonical hazard does not invalidate the correctness or even usefulness of the outputs.
For instance, the STPA Handbook~\cite{STPA_Handbook,leveson_saferworld} itself (and the prompt passed into the LLM) suggests that the loss scenarios need not be linked to a specific UCA.
Indeed, loss scenarios that are purely linked to a hazard make up a valid subtype of loss scenarios.
The presence of these outputs also suggests that the LLM was able to balance the number of loss scenario (sub)types.

In general, while we believe our case study of UCA and loss scenario syntheses is still preliminary, it shows great accuracy and usefulness to the requirement engineer.

\subsection{Limitations}

It is possible that our case studies, being publicly available along with all the other existing literature, have been included in the LLMs' training data and hence the generated artifacts were impacted by the data leakages.
With that said, we have qualitatively found that the LLM outputs are largely novel compared to the few-shot examples provided from the handbook~\cite{STPA_Handbook}.

Our case study is very small and limited to the examples in the STPA Handbook.
It is also possible that our annotations have incorrectly marked the synthesized UCAs and loss scenarios.
A much further study of our framework with various prompting strategies is needed to make the synthesis much more accurate.
Furthermore, in future studies including real exemplars outside the publicly available ones will be useful to assess the effectiveness of our approach.

We also understand the ethical limitations of leaving safety-critical work to LLMs as opposed to humans.
With that said, there exists ample previous literature that makes the case for using LLMs for safety engineering and finds these methodologies worthwhile to explore.
And, as our experimental result suggests, our framework has the potential to speed up traditional tasks associated with the STPA hazard analysis technique.
Note that, in the workflow model we propose in Figure~\ref{fig:process}, we do not replace the human with an LLM.
Instead, we keep a human-in-the-loop who guides the LLM generation (through few-shot examples) and verifies the LLM outputs based on correctness and usefulness.

\subsection{Future Work}

Qi et al.~\cite{era_LLM_STPA} propose to apply LLMs to other hazard analysis techniques such as FMEA and HAZOP, but we believe that STPA's more systematic nature is more conducive to LLM prompt generation.

Our experiment so far assumed only a static analysis of system hazards, but real-world systems evolve, introducing unforeseen risks.
Future work could explore AI-driven real-time safety monitoring, analyzing live system logs to detect emerging hazards and UCAs.
Given the high-intelligence of software defined vehicles, we can imagine an onboard STPA framework on automobiles.
This can be especially effective when all the STPA artifacts are explicitly linked to the software of the vehicle so that failures within the software or without (that relate to the software) would trigger certain requirements, alert the driver, or gracefully degrade certain functionalities.
Key challenges for the implementation of this feature would include automatically filtering incorrect outputs and duplicates, all while integrating adaptive safety frameworks that are capable of updating STPA artifacts dynamically in real-time.

Another possible research focus is automating risk assessments with LLMs.
Safety-critical insights often exist in unstructured data such as incident reports, audit logs, and operator feedback.
Applying NLP to the sources themselves may potentially reveal latent hazards, and link real-world failures to existing STPA artifacts.
In this case, we could extract relevant failure patterns, prioritize hazards, and develop domain-specific NLP models for different industries.

It may also be interesting to research the effectiveness of AI-enhanced STPA by analyzing how engineers actually interpret and validate the AI-generated artifacts.
Future work would examine trust, adoption, and bias in AI-assisted safety analysis.
This research could improve how engineers interact with AI-generated UCAs and loss scenarios, and provide insights on how the results lead to informed decision-making.

Purely from a performance perspective, the impact of different prompting techniques outside of the few-shot prompting we use (e.g., zero-shot, chain-of-thought, and more) would be interesting to look at, especially if they offer higher accuracy or response quality.

In our framework, we give several example system definitions from prior works~\cite{STPA_Handbook, autohold_STPA_exemplar}.
While the case studies of STPA for different systems are available as reports, publications, or theses, the artifacts generated from the analyses are often encoded in free text within PDF documents.
Therefore, programmatically analyzing them presents a time-consuming endeavor for interested researchers in the field.
It may be interesting to consolidate all open and publicly-available STPA applications in a single data source in Python (as we propose) for others to easily download and experiment, with our framework as a backbone.

\section{Conclusion}

In this paper, we introduce our completely free and open-source software library that offers safety and requirement engineers a framework to manage STPA and explicitly link its artifacts to the software, along with several LLM integrations to experiment with, designed to save time and effort put in by the human engineers during hazard analysis.
In addition to the classes from which definitions can be defined and parsers to extract information from control structure diagrams, our LLM-integrated tools support the syntheses of UCAs and loss scenarios given an STPA context and automated linkages.
In our experimentation, we have identified that an LLM can indeed be effectively used for the tasks we propose in this paper as they achieve acceptable accuracy and their response quality is mostly correct and useful, as identified by a human expert.
Possible future avenues of research would be to enhance this library further with more (LLM-powered) features, as well as an increased number of exemplars to demonstrate the usage of our software library.

\bibliographystyle{IEEEtran}
\bibliography{main}

\vfill

\end{document}